\documentclass{birkmult}
\usepackage{amsthm}
\usepackage{amssymb}
\usepackage{amsmath}
\usepackage{eucal}
\usepackage{mathrsfs}
\usepackage[all]{xy}
\usepackage{graphicx}
\usepackage{bbm}

\newcommand{\g}{\mathrm{g}} 
\newcommand{\Al}{\mathscr{A}} 

\newcommand{\ide}{\mathbf{1}}

\newcommand{\Loc}{\mathrm{\mathbf{Loc}}} 
\newcommand{\Obs}{\mathrm{\mathbf{Obs}}} 
\newcommand{\Dl}{\mathscr{D}} 
\newcommand{\El}{\mathscr{E}}  %

\newcommand{\supp}{\mathrm{supp}} 
\newcommand{\ord}[1]{\mathrm{ord}{(#1)}} 

  \theoremstyle{plain}

  \theoremstyle{definition}

\theoremstyle{definition}
  
\begin{document}
%
\firstpage{113}
\Copyrightyear{2006}
%
%
%
%
%
%
%
%
\title[Background Independent Formulation of Quantum Gravity]
{Towards a Background Independent Formulation of Perturbative Quantum Gravity}
\author{Romeo Brunetti}
\address{%
II Institute f\"ur Theoretische Physik\\
Universit\"at Hamburg\\
Luruper Chaussee 149, D-22761 Hamburg\\ 
Germany}
\email{romeo.brunetti@desy.de}
%
\author{Klaus Fredenhagen}
\address{%
II Institute f\"ur Theoretische Physik\\
Universit\"at Hamburg\\
Luruper Chaussee 149, D-22761 Hamburg\\ 
Germany}
\email{klaus.fredenhagen@desy.de}

\begin{abstract}
The recent formulation of locally covariant quantum field theory may open the way towards a background independent 
perturbative formulation of Quantum Gravity.
\end{abstract}

\maketitle

\section{Problems of perturbative Quantum Gravity}

In quantum field theory the fields are defined as operator-valued distributions on a given spacetime, and many
of their properties, in particular the 
commutativity at spacelike separated points, depend in a crucial way on properties of the background. In a 
perturbative approach to quantum gravity,
 one decomposes the metric $g_{\mu\nu}$ into a background metric $\eta_{\mu\nu}$ and a quantum field 
$h_{\mu\nu}$, which is treated according
 to standard methods in perturbation theory. The (up to now observed) effects of this quantum field are 
very small, hence a perturbative approach seems
 to be appropriate. There are, however,  several obstructions which raise doubts on the validity of the perturbative approach:
 \begin{enumerate}
\item The arising quantum field theory is nonrenormalizable \cite{GS}. Hence, infinitely many counterterms occur in the 
process of renormalization, and it is unclear how the arising ambiguities can be fixed \cite{W}.
\item The perturbatively constructed theory depends on the choice of the background. It is unlikely that a 
perturbative formulation can describe a drastic change of the background.
\end{enumerate}
The two main approaches to quantum gravity try to cope with these difficulties in different ways. 
String theory, in a first attempt, accepts the choice of a fixed background,
and aims at a more general theory where the perturbation series is finite in every order. Loop quantum gravity, 
on the other hand, uses a background free formulation
where the degrees of freedom of gravity are directly quantized. Problems with renormalizability do not occur, 
but it seems to be difficult to check whether such  theories describe the world as we see it.

Instead of following these routes one may take a more conservative approach and study first the influence 
of classical gravitational fields on quantum fields. Because of the weakness
of gravitational forces this approximation is expected to have a huge range of validity. Surprisingly,  this seemingly 
modest approach leads to many conceptual insight, and it may even lead
to a new approach to quantum gravity itself \cite{Lisbon2003}.

In this paper we want to review the recently developped new formulation of quantum field theory on 
curved spacetimes, which satisfies 
the conditions of general covariance \cite{BFV,HW1,HW2,V}. We will show that the arising structure has great 
similarities with Segal's concept of topological quantum field theories \cite{S} and its generalization to 
Riemannian spaces. It may be considered as a Lorentzian version of this approach. It is gratifying that the 
axiom of local commutativity is implied in this framework by the tensorial structure of the theory, while 
the time slice axiom (i.e., a form of dynamics) is related to cobordisms. 

It is remarkable that the new structures emerged from the (finally successful) attempt 
to construct interacting theories in the sense of renormalized perturbation theory. 

Up to now the complete proofs apply only to scalar field theories. 
The extension to gauge theories requires the control of BRST invariance. Preliminary steps in this direction 
have been performed \cite{DF1,DF2,DF3}, and no obstruction is visible.  

More or less, the same construction then should apply to 
quantum gravity, treated in a background formulation.
The leading idea is that background independence can be reached from a background dependent formulation provided 
the change of background amounts to a symmetry of the theory.

\section{Locally covariant quantum field theory}
We adopt the point of view \cite{BFV} of algebraic quantum field theory and identify physical systems with 
$*$-algebras with unit (if possible, C$^{*}$-algebras) and subsystems with 
subalgebras sharing the same unit. In quantum field theory the subsystems can be associated to 
spacetime regions. Every such region may be considered as a spacetime in its 
own right, in particular it may be embedded into different spacetimes. It is crucial that the algebra 
of the region does not depend on the way it is embedded into a larger spacetime.
For instance, in a Schwartzschild spacetime the physics outside the horizon should not depend on a 
possible extension to a Kruskal spacetime.

We formulate our requirements in form of five axioms:

\begin{enumerate}
\item Systems: To each time oriented globally hyperbolic spacetime $M$ we associate a unital $*$-algebra $\Al(M)$.
\item Subsystems: Let $\chi : M\rightarrow N$ be an isometric causality preserving embedding of globally 
hyperbolic spacetimes. Then, there exists a uniquely
defined (injective) $*$-homomorphism $\alpha_{\chi} : \Al(M)\rightarrow\Al(N)$.
\item Covariance: If $\chi: M_{1}\rightarrow M_{2}$ and $\chi^{\prime}:M_{2}\rightarrow M_{3}$ are embeddings 
as above, then $\alpha_{\chi\chi^{\prime}}
= \alpha_{\chi}\alpha_{\chi^{\prime}}$.
\item Causality: If $\chi_{1}:M_{1}\rightarrow M$ and $\chi_{2}: M_{2}\rightarrow M$ are embeddings as above, 
such that $\chi_{1}(M_{1})$ and $\chi_{2}(M_{2})$
cannot be connected by a causal curve in $M$, then
\[
\alpha_{\chi_{1}}(\Al(M_{1}))\vee \alpha_{\chi_{2}}(\Al(M_{2})) \simeq \alpha_{\chi_{1}}(\Al(M_{1}))\otimes 
\alpha_{\chi_{2}}(\Al(M_{2}))
\]
where $\vee$ indicates the generated subalgebra of $\Al(M)$.
\item Dynamics: Let $\chi :M \rightarrow N$ be an embedding as above such that $\chi(M)$ contains a Cauchy 
surface of $N$. Then $\alpha_{\chi}(\Al(M))=\Al(N)$.
\end{enumerate}

The axioms above describe a functor $\Al$ from the category $\Loc$ (the localization category) whose objects 
are time-oriented globally hyperbolic spacetimes
and whose arrows are the causal isometric embeddings, to the category $\Obs$ (the observables category) 
whose objects are unital $*$-algebras and whose arrows are (injective) $*$-homomorphisms. 

\noindent Axiom 1 is similar to the usual axiom in local quantum theories on a fixed background, where 
the arrow has specific (un)bounded regions on that background as a domain. 
Here, it is imperative to quantize \emph{simultaneously} on all globally hyperbolic spacetimes (of the given type). 

\noindent Axiom 2 may be pictured in the form
\[
\xymatrix{
 M \ar[d]_\Al \ar[r]^{\chi} & N \ar[d]^\Al\\
\Al(M)  \ar[r]^{\alpha_\chi} &\Al(N)}
\]
where $\alpha_{\chi}\doteq\Al\chi$.

\noindent Axiom 3 says that the functor $\Al$ is covariant. 

\begin{figure}[htbp]
\centering
\includegraphics[width=10truecm]{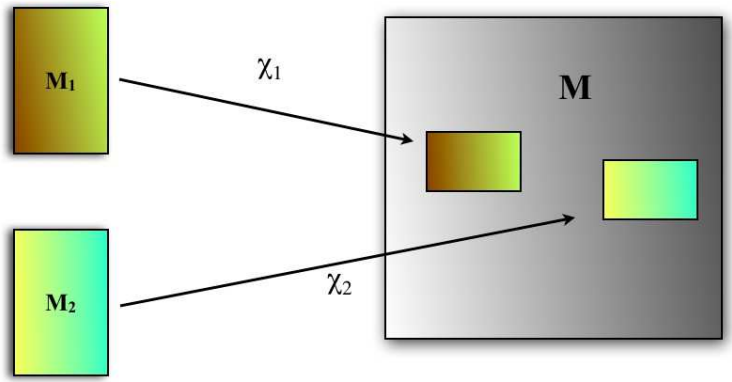}
\caption{Causal Embedding}
\label{fig:embedding}
\end{figure}

\noindent Axiom 4 may be reformulated in terms of a tensor structure. Namely, require for disjoint unions,
\[
\Al(M_{1}\amalg M_{2}) = \Al(M_{1})\otimes\Al(M_{2})\ , \qquad \Al(\emptyset)=\mathbb{C}\ ,
\]
with $\chi_{i}: M_{i}\rightarrow M, i=1,2$ for which $\alpha_{\chi_{1}\amalg\chi_{2}}=
\alpha_{\chi_{1}}\otimes\alpha_{\chi_{2}}$.
Let $\chi$ be a causal embedding of $M_{1}\amalg M_{2}$ into $M$. 
Then 
$\chi(M_{1})$ and $\chi(M_{2})$ are spacelike separated, hence with $i_{k}: M_{k}\rightarrow M_{1}\amalg M_{2}$, and with
$\chi_{k}=\chi\circ i_{k}$ (see fig.\ref{fig:embedding}), we see that $\alpha_{\chi}(\Al(M_{1})\otimes\Al(M_{2}))$ is equal to the algebra generated by
$\alpha_{\chi_{1}}(\Al(M_{1}))$ and $\alpha_{\chi_{2}}(\Al(M_{2}))$, hence the causality axiom is satisfied. In short, 
the functor $\Al$ is promoted to a \emph{tensor} functor. This is very reminiscent of G. Segal's approach \cite{S}.

\begin{figure}[htbp]
\centering
\includegraphics[width=10truecm]{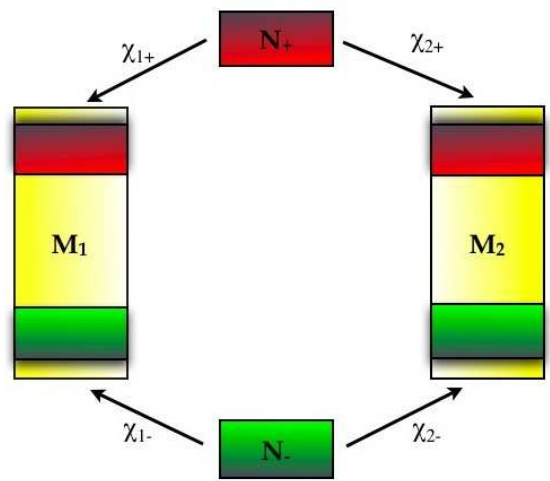}
\caption{Evolution}
\label{fig:evolution}
\end{figure}

\noindent Axiom 5 may be interpreted as a description of motion of a system from one Cauchy surface to another. Namely, 
let $N_{+}$ and $N_{-}$ be 
two spacetimes that embed into two other spacetimes $M_{1}$ and $M_{2}$ around Cauchy surfaces, via 
causal embeddings given by $\chi_{k, \pm}, k=1,2$. Figure \ref{fig:evolution} gives a hint.

Then $\beta =\alpha_{\chi_{1+}}\alpha_{\chi_{2+}}^{-1}\alpha_{\chi_{2-}}\alpha_{\chi_{1-}}^{-1}$ is an automorphism
of $\Al(M_{1})$. One may say that in case $M_{1}$ and $M_{2}$ are equal as topological manifold
but their metrics differ by a (compactly supported) symmetric tensor $h_{\mu\nu}$ with $\supp h\cap J^{+}(N_{+})\cap J^{-}(N_{-})=\emptyset$, 
the automorphism depends 
only on the spacetime between the two Cauchy surfaces, hence in particular, on the tensor $h$. It can then be shown
that
\[
\Theta_{\mu,\nu}(x) \doteq \frac{\delta\beta_h}{\delta h_{\mu,\nu}(x)}|_{h=0}
\]
is a derivation valued distribution which is covariantly conserved, an effect of the diffeomorphism invariance 
of the automorphism $\beta_h$,
and may be interpreted as the commutator with the energy-momentum tensor. Indeed, in the theory of the free 
scalar field this has been explicitely verified \cite{BFV}, and it remains true in perturbatively constructed 
interacting theories \cite{HW4}.

The structure described may also be understood as a version of cobordism. Namely, one may associate to a Cauchy 
surface $\Sigma$ of the globally hyperbolic spacetime $M$, the algebra
\[
\Al(\Sigma)\doteq \varprojlim_{N\supset\Sigma} \Al(N)\ ,
\]
where the inverse limit extends over the globally hyperbolic neighborhoods $N$ of $\Sigma$.
Clearly, $\Al(\Sigma)$ depends only on the germ of $\Sigma$ as a submanifold of $M$. The elements of
$\Al(\Sigma)$ are germs of families $(A_N)_{N\supset\Sigma}$ with the coherence condition 
$\alpha_{N_1 N_2}(A_{N_2})=
A_{N_1}$, where $\alpha_{N_1 N_2}$ is the homomorphism associated to the inclusion $N_2\subset N_1$.

We then define a homomorphism
\[
\alpha_{M\Sigma}: \Al(\Sigma)\rightarrow\Al(M)
\]
by $\alpha_{M\Sigma}(A)\doteq\alpha_{M N}(A_N)$, $N\supset\Sigma$, where the r.h.s. is independent of the chioce of the neighborhood $N$. By the time slice axiom, $\alpha_{M\Sigma}$ is invertible, hence for a choice of two
Cauchy hypersurfaces $\Sigma_1$, $\Sigma_2$ of $M$ we find a homomorphism
\[
\alpha_{\Sigma_1\Sigma_2}^{M} :\Al(\Sigma_2)\rightarrow\Al(\Sigma_1)\ ,
\]
with $\alpha_{\Sigma_1\Sigma_2}^{M}\doteq\alpha_{M\Sigma_1}^{-1}\alpha_{M\Sigma_2}$. We may interepret $\Sigma_1$
and $\Sigma_2$ as past and future boundaries, respectively, of $M$ and obtain for any spacetime $M$ connecting $\Sigma_1$ and $\Sigma_2$ a homomorphism of the corresponding algebras.

\section{Locally covariant fields}
One problem with a theory on a generic spacetime is that it is not clear what it means to do the 
same experiment at different spacetime points.
In quantum theory we are however forced to repeat the experiments in order to obtain a probability 
distribution. In a spacetime with a large symmetry group
one may use these symmetries to compare measurements on different spacetime points.
In the framework of locally covariant quantum field theory, as described above, quantum fields can serve 
as means for comparison of observables at different points. 
Namely, we define a locally covariant quantum field $A$ as a family of algebra valued 
distributions $(A_{M})$ indexed by
the objects $M\in\Loc$ which satisfy the following covariance condition
\[
\alpha_{\chi}(A_{M}(x)) = A_{N}(\chi(x))\ , \qquad \chi: M\rightarrow N\ .
\]
More formally, one may define a locally covariant quantum field as 
a natural transformation from the functor $\Dl$, that associates to each manifold its
test-function space, to the functor $\Al$. Actually, we pinpoint that fields, 
differently from the traditional point of view, are now objects as fundamental as observables. Not only, 
they might even be more fundamental in cases like that of quantum gravity
where \emph{local} observables would be difficult to find (if they exist at all). 

Let us look at an example. We define the theory of a real Klein-Gordon field in terms of the algebras 
$\Al_{0}(M)$ which are generated
by elements $\varphi_{M}(f)$, $f\in\Dl(M)$, satisfying the relations
\begin{enumerate}
\item[$(i)$]  $f\rightarrow \varphi_{M}(f)$ is linear;
\item[$(ii)$] $\varphi_{M}(f)^{\ast}=\varphi_{M}(\overline{f})$;
\item[$(iii)$] $\varphi_{M}((\square_{M} +m^{2})f)=0$;
\item[$(iv)$] $[\varphi_{M}(f),\varphi_{M}(g)]=i (f,\Delta_{M}g)\ide$
\end{enumerate}
where $\Delta_{M}=\Delta_{M}^{\mathrm{ret}}-\Delta_{M}^{\mathrm{adv}}$ with the retarded and advanced propagators
of the Klein-Gordon operator, respectively.
The homomorphism $\alpha_{\chi}$, $\chi: M\rightarrow N$, is induced by
\[
\alpha_{\chi}(\varphi_{M}(f)) = \varphi_{N}(\chi_{\ast}(f))
\]
where $\chi_{\ast}f$ is the push-forward of the test function $f$.
We see immediately that $\varphi=(\varphi_{M})$ is a locally covariant quantum field associated to the functor
$\Al_0$.

To find also other locally covariant fields 
it is convenient to enlarge the previuosly constructed algebra by neglecting the field equation
 $(iii)$. Notice that the new functor $\Al_{00}$ does no longer satisfy the time slice axiom. We then 
introduce localized polynomial
functionals $\mathscr{F}(M)$ on the space of 
classical field configurations $\phi\in C^{\infty}(M)$, 
\[
\mathscr{F}(M)\ni F(\phi) = \sum_{n=0}^{\ord{F}} \langle f_{n},\phi^{\otimes n}\rangle\ ,\qquad \phi\in C^{\infty}(M)
\]
 where $f_{n}\in\El^{\prime}_{\Gamma_n}(M^{n})$, i.e., $f_n$ is a distribution of compact support whose wave front set 
$\mathrm{WF}(f_{n})\subset\Gamma_n$ and $\Gamma_n\cap\{(x,k)\in T^{\ast}M^{n}, 
k\in \overline{V}_{+}^{n}\cup\overline{V}_{-}^{n}\}=\emptyset$.
 We choose a decomposition of $\Delta_{M}$, 
 $\Delta_{M}(x,y)=H(x,y) - H(y,x)$, 
 such that 
$\mathrm{WF}(H)=
\{(x,k)\in \mathrm{WF}(\Delta_{M}), k\in\overline{V}_{+}\times
 \overline{V}_{-}\}$. (This is a microlocal version of the decomposition into positive and negative frequencies,
for explanations see, for instance, \cite{BFK}.) 
Then, we define a product on $\mathscr{F}(M)$ by
 
 \[
 F\ast_{H}G = \sum_{n}\frac{1}{n!}\ \left\langle \frac{\delta^{n} F}{\delta\phi^{n}}
\otimes\frac{\delta^{n} G}{\delta\phi^{n}}, H^{\otimes n}\right\rangle\ ,
 \]
 which makes $\mathscr{F}(M)$ to an associative algebra $(\mathscr{F}(M),\ast_H)$.

But $H$ is not unique. If we change $H$ to $H^\prime=H+w$, $w\in C^{\infty}_{\text{symm}}(M^2)$ (since the difference between two 
$H$'s is always a smooth symmetric function), we find
 
 \[
 \gamma_{w}(F\ast_{H}G) = \gamma_{w}(F)\ast_{H'}\gamma_{w}(G)
 \]
 with
 
 \[
 \gamma_{w}(F) = \sum_{n}\frac{1}{2^{n}n!}\left\langle \frac{\delta^{2n}F}{\delta\phi^{2n}}, w^{\otimes n}\right\rangle \ .
 \]
The algebra $\Al_{00}(M)$ may be embedded into $(\mathscr{F}(M),\ast_H)$ by
 
 \[
 \alpha_{H}(\varphi_{M}(f)) = \langle f , \phi \rangle\ ,
 \]
 
hence, in particular
 \[
 \alpha_{H}(\varphi_{M}(f)\varphi_{M}(g)) = \langle f\otimes g, \phi^{\otimes 2}\rangle + \langle f, Hg\rangle\ .
 \]
 
 Then $\alpha_H(\Al_{00}(M))$ is a subalgebra of $(\mathscr{F}(M),\ast_H)$ with coefficients $f_n\in\mathscr{D}(M^n)$. 
Since the space $\mathscr{F}(M)$ is uniquely determined by the coefficients, we may equip it with the 
inductive topology of the direct sum of the spaces $\mathscr{E}^\prime_{\Gamma_n}(M^n)$. Since $\Dl(M^n)$ is dense
in $\El^{\prime}_{\Gamma_n}(M^n)$, $\alpha_H(\Al_{00}(M))$ is dense in $\mathscr{F}(M)$. We may now equip 
the algebra $\Al_{00}(M)$ with the initial topologies of $\alpha_{H}$. Since $\gamma_w$ is a homeomorphism, it 
turns out that all induced topologies coincide, and the completion is identified with our seeked algebra $\Al(M)$.

This algebra contains, besides the usual fields, also their normal ordered products.

Now, for the sake of constructing interacting quantum fields one may use the following steps:
\begin{enumerate}
\item Construction of locally covariant Wick polynomials; this turns out to the solution of a cohomological 
problem \cite{BFV} (see also \cite{HW2}) which can be solved in terms of an explicit Hadamard parametrix of 
the Klein Gordon equation.
\item Construction of locally covariant retarded and time ordered products; this requires the generalization 
of the Epstein-Glaser renormalization scheme to curved spacetime \cite{BF} and again a solution of a cohomological 
problem in order to be able to impose the same renormalization conditions on every point of a given spacetime and 
even on different spacetimes. 
\item Construction of the algebras of interacting fields, together with a family of locally covariant fields 
\cite{BF,DF2,HW1,HW2}.
\end{enumerate} 

We refrain from giving details of these steps and refer to the original publications \cite{BF,BFV,DF2,HW1,HW2,HW3,HW4}.

\section{Quantization of the background}
As usual, we view the metric as a background plus a fluctuation, namely,
\[
\g_{\mu\nu}=\eta_{\mu\nu} + h_{\mu\nu}
\]

and we look at the fluctuation as a quantum field. Note that differently from other approaches 
the background metric does not need to be Minkowskian, we only restrict to backgrounds complying with
the requirements of the previous sections.

So we have a quantum field $h$ that propagates via the linearized Einstein equations on a fixed background $\eta$.

One may now proceed by the general strategy for constructing gauge theories by the BRST method.
It is crucial that this method can be adapted to localized interactions, as was done in \cite{DF1,DF3}.
Furthermore, the freedom in renormalization can be used to arrive at quantized metric and curvature fields satisfying Einstein's equation. 

The condition of background independence may be formulated as the condition that the automorphism 
$\beta_\kappa$ describing the relative Cauchy evolution corresponding to a change of the background between  two Cauchy surfaces must be trivial.  In perturbation theory, it is sufficient to check the infinitesimal version of this condition. This amounts to the equation
\[
\frac{\delta\beta_\kappa}{\delta\kappa_{\mu\nu}(x)}=0\ .
\]
In contrast to the situation for an unquantized background metric the left hand side involves in 
addition to the energy momentum tensor also the Einstein tensor. Hence the validity of Einstein's
equation for the quantized fields should imply background independence.

There remain, of course, several open questions. First of all, the details of the proposal above have to be elaborated, and in particular the question of BRST invariance has to be checked. A possible obstruction could be that locally the cohomology of the BRST operator is trivial, corresponding to the absense of local observables in quantum gravity. 
Another problem is the fact that the theory is not renormalizable by power counting. Thus the theory will have the status of an effective theory. Nevertheless, due to the expected smallness of higher order counter terms  the theory should still have predictive power.

%
\newpage
\tableofcontents
\newpage
\printindex
\end{document}